\documentstyle[a4,11pt]{article} 

\input amssym.def       
\input amssym.tex

\def\double{\Bbb}

\def\cc{{\double C}}     
       
\def\zz{{\double Z}}
\def\qq{{\double Q}}

\newtheorem{theorem}{Theorem}

\newtheorem{corollary}[theorem]{Corollary}
\newtheorem{definition}[theorem]{Definition}

\newtheorem{remark}[theorem]{Remark}
\def\res{\mathop{\mathrm{Res}}\limits_{z=0}}
\def\cp{>\!\!\!\lhd}

\newcommand{\be}{\begin{equation}}
\newcommand{\ee}{\end{equation}}
\newcommand{\beq}{\begin{eqnarray}}
\newcommand{\eeq}{\end{eqnarray}}

\newcommand{\al}{\alpha}

\newcommand{\Vc}{{\cal V}}
\newcommand{\Lc}{{\cal L}}
\newcommand{\non}{\nonumber}

\newcommand{\Ind}{\mbox{Ind}}
\newcommand{\ch}{\mbox{ch}}

\newcommand{\Tr}{\mbox{Tr}}
\newcommand{\tr}{\mbox{tr}}
\newcommand{\Ac}{{\cal A}}

\newcommand{\te}{\theta}
\newcommand{\Te}{\Theta}

\begin{document}

\begin{center}

{\large BRS COHOMOLOGY AND THE CHERN CHARACTER IN NON-COMMUTATIVE GEOMETRY\\}
\vskip 1cm
{\bf Denis PERROT\footnotemark[1]}
\vskip 0.5cm
Centre de Physique Th\'eorique, CNRS-Luminy,\\ Case 907, 
F-13288 Marseille cedex 9, France \\[2mm]
{\tt perrot@cpt.univ-mrs.fr}
\end{center}
\vskip 1cm
{\bf Abstract:} We establish a general local formula computing the topological anomaly of gauge theories in the framework of non-commutative geometry.

\vskip 1cm

\noindent {\bf MSC91:} 19D55, 81T13, 81T50\\

\footnotetext[1]{Allocataire de recherche MRT.}

\noindent {\bf I. Introduction}\\

The relationship between topological anomalies in Quantum Field Theory and classical index theorems is an old subject. The mathematical understanding was investigated by Atiyah and Singer in \cite{AS}, and further related to Bismut's local index formula for families in \cite{BF,F} (see e.g. \cite{AG} for a more physical presentation). In this paper, we shall describe quantum anomalies from the point of view of non-commutative geometry \cite{C2}. The latter deals with a large class of ``spaces'' (including pathological ones such as foliations, etc.) using the powerful tools of functional analysis. The advantage we can take of this theory in QFT is obvious. In our present work, it turns out that anomalies, and more generaly BRS cohomology, are just the pairing of odd cyclic cohomology with algebraic $K_1$-groups \cite{C1,C2}. Although these tools are well-known and well-used by mathematicians, it is still unclear whether the physics literature has assimilated it. We will try to fill this gap below.\\

According to Connes \cite{C2}, a non-commutative space is described by a spectral triple $(\Ac,H,D)$, where $\Ac$ is an involutive algebra of operators on the Hilbert space $H$, and $D$ is a selfadjoint unbounded operator. $D$ carries a nontrivial homological information through its Chern character in the cyclic cohomology of $\Ac$ \cite{C2}. In our field-theoretic interpretation, $H$ is a space of matter fields (e.g. fermions), and $D$ is a Dirac operator. We also introduce a Lie group $G\subset \Ac$, which plays the role of gauge transformations. Now BRS cocycles are obtained by transgression of the Chern character of an index bundle, constructed from a $G$-equivariant family of Dirac operators. The Chern character of this bundle is directly related to that of $D$.\\
The natural receptacle for the above construction should be Kasparov's
bivariant $K$-theory \cite{Sk}. For some convenience, we give a direct proof of the needed cohomological formula, using only the pairing of cyclic cohomology with $K$-theory \cite{C1}. The final step is to use the local index formula of Connes-Moscovici \cite{CM95} in order to get explicit expressions of consistent anomalies.\\
We end the paper by mentioning two examples. The first one concerns
the comp
utation of the Yang-Mills anomaly on a Riemannian spin manifold, which makes only use of the underlying classical (commutative) geometry. The second example is provided by the gravitational anomaly. It is much more involved because of the natural non-commutative structure encoded in diffeomorphisms. This constitutes a relevant illustration of non-commutative index theorems in QFT.\\
\vskip 1cm

\noindent {\bf II. Conventions}\\

We begin with some ingredients describing our non-commutative space. Let $\Ac$ be a unital *-algebra represented in the algebra $\Lc (H)$ of bounded linear operators on a separable Hilbert space $H$, and $D$ a densely defined unbounded self-adjoint operator in $H$. The spectral triple $(\Ac,H,D)$ is supposed to satisfy the following properties \cite{C1}:\\
i) $\exists\ p\in [1,\infty[$ such that $(1+D^2)^{-1}\in \Lc^{p/2}(H)$,\\
ii) $\forall\, a\in\Ac$, $[D,a]$ is densely defined and extends to a bounded operator in $H$,\\
iii) there is a self-adjoint involution $\gamma\in \Lc(H)$ such that $\gamma D=-D\gamma$, $\gamma a=a\gamma\quad \forall a\in\Ac$;\\
that is, $(H,D)$ is an even $p$-summable unbounded Fredholm module over $\Ac$. We can choose the following matricial representation in the $\zz_2$-graded space $H=H^+ \oplus H^-$:
\be
D=\left( \begin{array}{cc}
          0 & D^- \\
          D^+ & 0 \\
     \end{array} \right) \qquad
\gamma=\left( \begin{array}{cc}
          1 & 0 \\
          0 & -1 \\
     \end{array} \right)\ .
\ee

Consider now a set of ``gauge potentials'': it is an affine subspace $\Vc\subset \Lc(H)$ of odd self-adjoint bounded operators. For any $A\in \Vc$, $D_A\equiv D+A$ is a bounded perturbation of $D$, and $(H,D_A)$ is a $p$-summable Fredholm module homotopic to $(H,D)$ \cite{CM93}. \\
Let $G\subset \Ac$ be a Lie group of unitary elements of $\Ac$. We assume that the natural action of $G$,
\beq
D_A &\rightarrow& g^{-1}D_A g\ ,\qquad\qquad g\in G\ ,\non\\
A&\rightarrow& g^{-1}Ag + g^{-1}[D,g]\ ,
\eeq
on $\Vc$ is free. This turns $\Vc$ into a principal $G$-bundle over the space $\Vc/G$, which we assume to be paracompact. These properties imply in particular that $\Vc/G$ is a specific realization of the classifying space $BG$, and that $\Vc$ is bundle-homotopic to the universal bundle $EG$ over $BG$.\\

We describe now the index bundle associated to the family $(H,D_A)$. Let $X$ be a compact subset of $\Vc/G$. The equivariant family of finite-dimensional spaces $\ker D_A^+\subset H^+$ and $\ker D_A^-\subset H^-$ defines a virtual bundle over $X$, the so-called index bundle 
\be
\Ind(H,D) = \ker D_A^+ -\ker D_A^-\ \in K^0(X)\ .
\ee
Our goal is to compute its Chern character in rational cohomology. We shall represent it in de Rham cohomology as follows. Let $M$ be a compact smooth manifold, and $f:M\rightarrow X$ a continuous map. The bundle $\Vc$ pulls-back on a bundle $\tilde{M}$ over $M$. Choose a connection one-form $\te$ on $\tilde{M}$, and define the Quillen superconnection \cite{Q}
\be
\nabla = d+\te+itD_A\ ,
\ee
where $d$ is the differential on $\tilde{M}$, and $t>0$ a real number. Then a result of Bismut \cite{BF} shows that the closed differential form $\Tr_s\, \exp \nabla^2$ represents the Chern character $\ch^*(\Ind)\in H^*(M;\qq)$ of the index bundle\footnotemark[1]. Here $\Tr_s=\Tr(\gamma.)$ denotes the supertrace of operators on the $\zz_2$-graded Hilbert space $H$.\\

\footnotetext[1]{Here and in the following we omit several $2\pi i$ factors. It causes no confusion since all the de Rham cohomology classes we are dealing with are understood to be rational.}

By a classical transgression \`a la Chern-Simons, one can use this Chern character to construct closed differential forms on the Lie group $G$. Consider the linear homotopy
\be
\nabla_u = d+itD+u(\te+itA)\ ,\qquad u\in [0,1]\ .
\ee
Then the Chern-Simons forms are obtained by the odd components in the differential form
\be
cs(H,D)= \int_0^1 du\,\Tr_s((\te+itA)e^{\nabla_u^2})\ .
\ee
When restricted to the fibres of $\tilde{M}$, the former are closed. These elements $\al_{2k+1}$ of de Rham cohomology $H_{dR}^{2k+1}(G;\cc)$ will be refered to as BRS cocyles. In particular, the topological anomaly is the element $\al_1$ of the first cohomology $H_{dR}^1(G;\cc)$.\\

At this point, one could be tempted to compute these classes by local formulas. For instance, using Duhamel's formula, and taking the finite part of $cs$ at the limit $t\rightarrow 0$, the result should be expressed in terms of residues of some zeta-functions, following the same lines as in \cite{CM95}. Since it is an easy transcription exercise, we shall not develop it further. Let us rather concentrate on the links between the Chern character of the index bundle and the one of the $p$-summable Fredholm module $(H,D)$ in periodic cyclic cohomology.\\

Such a link was already exhibited in \cite{CM86}. Assume for simplicity that $D$ is invertible (the general case can be treated as in \cite{CM93}). The Chern character $\ch_*(D)$ is represented, for any even integer $n>p-1$, by the following $n$-dimensional cyclic cocycle
\be
\tau_D(a_0,...,a_n)= {1\over 2}\left({n\over 2}\right)!\Tr_s(F[F,a_0]...[F,a_n])\ , \ F={D\over {|D|}}\ ,\ a_i\in\Ac\ .
\ee
Now for any $A\in\Vc$ such that $D_A$ is invertible, we define the $(n+1)$-dimensional differential form
\be
\Phi\tau_{D_A}= \tau_{D_A}(\te,...,\te)\ ,
\ee
where $\te$ is a connection form on $\Vc$. The restriction of $\Phi\tau_{D_A}$ to the fibres of $\Vc$ is closed and invariant under the right action of $G$. Next we obtain, by a straightforward adaptation of \cite{CM86}, the following

\begin{theorem}[\cite{CM86}]
Let $A\in\Vc$ be such that $D_A$ is invertible. Let $n>p-1$ be an even integer, $\tau_{D_A}$ the $n$-dimensional cyclic cocycle representing the Chern character of $(H,D_A)$, and $\al_{n+1}$ the BRS cocycle of degree $n+1$. Then the restriction of $\Phi\tau_{D_A}$ to the orbit of $A$ under the action of $G$ is\footnotemark[2]
\be
\Phi\tau_{D_A} = (n+1)!\,\al_{n+1}\quad in \ H_{dR}^{n+1}(G;\cc)\ .
\ee
\end{theorem}

{\it Hint of proof:} Consider the linear homotopy between the two superconnections $d+\te+itD_A$ and $d+itD_A$, for $t$ sufficiently large. The conclusion is a consequence of the contractibility of $\Vc$ and \cite{CM86} Thm. 9. \hfill $\Box$\\

\footnotetext[2]{This formula appears in \cite{CM86} with an erroneous $n!$ coefficient.}

For $n\le p-1$, the de Rham cohomology of $G$ cannot be represented by invariant forms. Thus we need another relation between $\ch^*(\Ind)$ and $\ch_*(H,D)$ to cover this case as well. This is done in the next sections.\\
\vskip 1cm

\noindent {\bf III. Cyclic cohomology and the index bundle}\\

Let $(\Ac,H,D)$ be a $p$-summable spectral triple as before (we don't have to assume that $D$ is invertible). In order to relate the Chern characters of both $\Ind(H,D)$ and the Fredholm module $(H,D)$, we need a more algebraic approach to the topological construction presented in the preceding section.\\

In order to give some flavour on what ought to happen in the case of Lie groups, we consider first the case where $G$ is discrete. Let $\bar{\Ac}$ be the norm closure of $\Ac$ in $\Lc(H)$. Take a compact subset $X\subset BG$ and form the $C^*$-algebra tensor product $C(X)\otimes\bar{\Ac}$, where $C(X)$ is the algebra of continuous functions on $X$. Now the $G$-principal bundle $\tilde{X}$ over $X$ can be explicitely described by the following idempotent $e\in M_n(C(X)\otimes\bar{\Ac})$ (``Mis\v{c}enko line bundle'', see \cite{C2} chap. II). Let $(U_i)_{i=1,n}$ be a finite open covering of $X$ such that $\tilde{X}$ is trivial over each $U_i$, and $g_{ij}:U_i\cap U_j\rightarrow G$ be the transition functions defining $\tilde{X}$. Take a partition of unity $(\rho_i)_{i=1,n}$ relative to the covering $(U_i)$, $\sum_i \rho_i^2 =1$. Then the component $e_{ij}$ of $e$ in the $n\times n$ matrices over $C(X)\otimes\bar{\Ac}$ is given by
\be
e_{ij}= \rho_i\rho_j g_{ij}\quad \in C(X)\otimes\bar{\Ac}\ .
\ee
The equivalence class of the idempotent $e$ defines an element $[e]$ of the Kasparov group $KK(\cc, C(X)\otimes\bar{\Ac})$. Also $(H,D)$ defines a class in $KK(\bar{\Ac},\cc)$, the set of homotopy classes of Fredholm modules over $\bar{\Ac}$. Now the Kasparov product \cite{Sk}
\be
\otimes_{\bar{\Ac}}:\ KK(\cc, C(X)\otimes\bar{\Ac})\times KK(\bar{\Ac},\cc)\rightarrow  KK(\cc, C(X))
\ee
enables one to give the following

\begin{definition}
The index bundle {\normalsize $\Ind(H,D)\in KK(\cc,C(X))$} over $X$ is given by the Kasparov product {\normalsize
\be
\Ind(H,D)= [e]\otimes_{\bar{\Ac}} (H,D) \label{ind}
\ee
} of the Mis\v{c}enko line bunble by the Fredholm module $(H,D)$.
\end{definition}

We now deal with Chern characters and a non-necessarily discrete Lie group $G$. Let $M$ be an even-dimensional oriented smooth manifold without boundary, and $M\rightarrow X$ a continuous map. We represent the Chern character of $\Ind(H,D)$ in the cohomology of the de Rham complex of differential forms $\Omega^*(C^{\infty}(M),d)$. On the other hand, the Chern character of $(H,D)$ is represented by a cyclic cocycle over $\Ac$, or equivalently \cite{C1}, by a closed graded trace on the enveloping differential algebra $\Omega^*(\Ac,\delta)$.\\
Finally, the (smooth) idempotent $e\in M_n(C^{\infty}(M)\otimes\Ac)$ has a Chern character constructed from a curvature two-form as in Chern-Weil theory. Consider the graded differential algebra $\Omega^*(C^{\infty}(M)\otimes\Ac,d+\delta)$. Then $e$ determines a canonical curvature
\be
\Te = e(d+\delta)e(d+\delta)e\quad \in \Omega^2(M_n(C^{\infty}(M)\otimes\Ac),d+\delta)\ ,
\ee
and its Chern character in $K_0$-theory reads
\be
\ch_0^*(e)= \sum_k {1\over{k!}}\tr_n\Te^k\quad \in \Omega^{even}(C^{\infty}(M)\otimes\Ac,d+\delta)\ .
\ee
Note that the natural map from $\Omega^*(C^{\infty}(M)\otimes\Ac,d+\delta)$ to the graded tensor product of differential algebras $\Omega^*(C^{\infty}(M),d)\hat{\otimes}\Omega^*(\Ac,\delta)$ yields a cup product 
\be
\cup: H_n(M;\cc)\otimes HC^m(\Ac)\rightarrow HC^{n+m}(C^{\infty}(M)\otimes\Ac)
\ee
between the de Rham homology of $M$ and the cyclic cohomology of $\Ac$, exactly as in \cite{C1}.\\

We can thus state the main theorem, which is the cohomological version of eq.(\ref{ind}):

\begin{theorem}
Let {\normalsize $\ch_*(D)$} be the cyclic cohomology Chern character of the $p$-summable Fredholm module $(H,D)$. We view it as a closed graded trace on $\Omega^*(\Ac,\delta)$. Then the Chern character of the index bundle is {\normalsize
\be
\ch^*(\Ind)=\langle \ch_0^*(e),\ch_*(D)\rangle \ \in H_{dR}^{even}(M;\cc)\ .\label{ch}
\ee }
\end{theorem}

{\it Proof:} Endow $M$ with a Riemannian metric. Let $(H_{\sigma},D_{\sigma})$ be the even Fredholm module associated to the signature operator $[\sigma]\in K_*(M)$ of $M$. Consider the graded tensor product of $(H_{\sigma},D_{\sigma})$ with $(H,D)$: it is the even module $(H_{\sigma}\hat{\otimes}H, D_{\sigma}\hat{\otimes}1+1\hat{\otimes}D)$, whose Chern character is the cup product of $\ch_*(D_{\sigma})\in H_*(M;\qq)$ with $\ch_*(D)\in HC^*(\Ac)$. Now the index of $P_e= e(D_{\sigma}\hat{\otimes}1+1\hat{\otimes}D)e$ is given by Connes' pairing 
\be
\mbox{index}P_e = \langle \ch_0^*(e), \ch_*(D_{\sigma})\cup\ch_*(D)\rangle\ .
\ee
Moreover $P_e$ describes also the signature operator twisted by the index bundle, so that the classical Atiyah-Singer theorem gives
\be
\mbox{index}P_e = \langle\ch^*(\Ind),\ch_*(D_{\sigma})\rangle\ .
\ee
Let $E$ be an arbitrary vector bundle over $M$. If we twist $P_e$ again with $E$, we are left with
\be
\langle\ch_0^*(e), (\ch^*(E)\cap \ch_*(D_{\sigma}))\cup\ch_*(D)\rangle = \langle\ch^*(\Ind),\ch^*(E)\cap\ch_*(D_{\sigma})\rangle\ .
\ee
The Chern isomorphism $\ch^*: K^*(M)\otimes\qq\rightarrow H^{even}(M;\qq)$ allows one to ``simplify'' both sides by $\ch^*(E)\cap\ch_*(D_{\sigma})\in H_*(M;\qq)$ and the conclusion follows. \hfill $\Box$\\

Eq.(\ref{ch}) can be transgressed to $H^{odd}(G;\cc)$ as usual. We introduce the map $g$ as the inclusion $g:G\hookrightarrow \Ac$. It is an invertible element of $C^{\infty}(G)\otimes\Ac$, which determines a canonical class $[g]$ in the algebraic $K$-group $K_1(C^{\infty}(G)\otimes\Ac)$. Its Chern character $\ch_1^*(g)$ can be expressed in the differential graded algebra $\Omega^*(C^{\infty}(G)\otimes\Ac, d+\delta)$ by means of the Maurer-Cartan form $\omega=g^{-1}(d+\delta)g$:
\be
\ch_1^*(g)= \sum_k (-)^k{{k!}\over{(2k+1)!}} \omega^{2k+1}\ \in \Omega^{odd}(C^{\infty}(G)\otimes\Ac, d+\delta)\ .
\ee
We deduce immediately 
\begin{theorem}
The BRS cocycles are given by the pairing {\normalsize
\be
\al= \langle\ch_1^*(g),\ch_*(D)\rangle \ \in H_{dR}^{odd}(G;\cc)\ ,\label{zozo}
\ee }
where the Chern character {\normalsize $\ch_*(D)$} is viewed as a closed graded trace on the enveloping differential algebra $\Omega^*(\Ac,\delta)$.
\end{theorem}
\hfill $\Box$\\

\begin{remark} \normalsize

Eq. (\ref{zozo}) is nothing else but the pairing between odd cyclic cohomology and $K_1$-groups \cite{C1}. Indeed, let $\Gamma\in H_{odd}(G,\zz)$ be an odd-dimensional integral cycle on $G$ considered as a smooth manifold. Then the evaluation of $\al$ on $\Gamma$ is the rational pairing of $[g]\in K_1(C^{\infty}(G)\otimes\Ac)$ with the odd cyclic cohomology class $\Gamma\cup\ch_*(D)\in HC^{odd}(C^{\infty}(G)\otimes\Ac)$:
\be
\langle\al, \Gamma\rangle = \langle [g], \Gamma\cup \ch_*(D)\rangle\ \in\qq\ .
\ee
\end{remark}

\begin{remark} \normalsize

The de Rham cohomology class of $\al$ does not depend on the dimension of the even cyclic cocycle representing $\ch_*(D)$, since $K$-theory pairs with {\it periodic} cyclic cohomology.
\end{remark}

\begin{remark} \normalsize

The above pairing is also available when the Fredholm module $(H,D)$ is only $\Te$-summable, provided its Chern character is expressed in entire cyclic cohomology \cite{C2}.

\end{remark} 

\vskip 1cm

\noindent {\bf IV. The local anomaly formula}\\

In this section we compute the topological anomaly explicitely by using the local formula of Connes and Moscovici for the Chern character of finitely summable spectral triples \cite{CM95}.\\

Consider a smooth map $g:S^1\rightarrow G$ from the circle to the Lie group $G$. It determines a canonical element $[g]\in K_1(C^{\infty}(S^1)\otimes\Ac)$. Our goal is therefore to evaluate the pull-back of the anomaly $g^*(\al_1)\in H_{dR}^1(S^1;\cc)$ on the fundamental class $[S^1]\in H_1(S^1;\zz)$, by the formula
\be
\int_{S^1}g^*(\al_1) = \langle [g], [S^1]\cup\ch_*(D)\rangle\ \in\zz \label{an}
\ee
This pairing is integral and can be expressed as the index of an operator. To see this, we introduce the spectral triple of the circle parametrized by $\te\in [0,2\pi]$:
\be
(C^{\infty}(S^1),L^2(S^1),-i\partial_{\te})\ .
\ee
Its Chern character in de Rham homology is just the fundamental class $[S^1]$. Thus $[S^1]\cup\ch_*(D)$ is the Chern character of the tensor product $(\Ac',H',D')=(C^{\infty}(S^1),L^2(S^1),-i\partial_{\te})\otimes(\Ac,H,D)$. The Dirac operator of this product is given by
\be
D'= -i\partial_{\te}\otimes\gamma + 1\otimes D= \left(
\begin{array}{cc}
-i\partial_{\te} & D^-\\
D^+ & i\partial_{\te} \end{array} \right) \ .
\ee
$(\Ac',H',D')$ is thus an odd triple, and we shall use the formula of \cite{CM95} to compute (\ref{an}). First, we have to make the following regularity hypothesis,
\be
a\ \mbox{and}\ [D',a]\in \cap\,\mbox{Dom}\delta^k\ ,\quad \forall a\in\Ac'
\ee
where $\delta$ is the derivation $\delta(P)=[|D'|,P]$ for any operator $P$. Furthermore, the dimension of $(\Ac',H',D')$ is described by the discret set $\Sigma\subset\cc$ of singularities of the zeta-functions
\be
\zeta_b(z)= \Tr(b|D'|^{-z})\ ,\quad z\in\cc\ ,
\ee
where $b$ is any element of the algebra generated by $\delta^k(a), \delta^k([D',a]), a\in\Ac'$. We suppose that $\zeta_b(z)$ extends holomorphically to $\cc\backslash\Sigma$.\\
Second, the Chern character of $(\Ac',H',D')$ is expressed in the $(b,B)$ bicomplex of $\Ac'$ \cite{C2}. Recall that the $b+B$ cohomology of cochains with finite length is isomorphic to periodic cyclic cohomology. The correspondence goes as follows. If $\psi$ is an $n$-dimensional cyclic cocycle, then the $(b+B)$-cochain
\be
\phi_n=(-)^{[n/2]} {1\over {n!}}\psi
\ee
is a $(b+B)$-cocycle whose cohomology class corresponds to the class of $\psi$ in periodic cyclic cohomology.\\

Then Connes and Moscovici show \cite{CM95} that the components $\phi_n$, $n$ odd (bounded by the summability degree), of the Chern character of $(\Ac',H',D')$ in the bicomplex are expressed by finite linear combinations of residues of the form
\be
\res \ z^q\Tr(a_0[D',a_1]^{(k_1)}...[D',a_n]^{(k_n)}|D'|^{-(2|k|+n+2z)})\ ,\quad a_i\in\Ac'\ ,
\ee
where $P^{(k)}$ is the $k^{th}$ iterated commutator of $P$ with $D'^2$, and $q,k_i$ are bounded positive integers.\\

Now the cocycle $\sum_n\phi_n$ pairs with the Chern character $\ch_1^*(g)$ of $[g]\in K_1(\Ac')$, whose components in cyclic homology read \cite{L}:
\be
\ch_1^{2k+1}(g)= (-)^k k!\, g^{-1}\otimes g\otimes ...g^{-1}\otimes g\ \in{\Ac'}^{\otimes 2(k+1)}\ .
\ee
We deduce finally

\begin{corollary}[local anomaly formula]
Let $g:S^1\rightarrow G$ be a smooth map. Then the integration of the anomaly $\al_1\in H_{dR}^1(G;\cc)$ on the circle is the integer
\be
\int_{S^1} g^*(\al_1) = \sum_k (-)^k k!\,\phi_{2k+1}(g^{-1},g,...,g^{-1},g)\ ,\label{toto}
\ee
where $\phi_n$, $n$ odd, are the components of the Chern character $\ch_*(-i\partial_{\te}\otimes\gamma+1\otimes D)$ in the $(b,B)$ bicomplex of $C^{\infty}(S^1)\otimes\Ac$.
\end{corollary}
\hfill $\Box$\\

\begin{remark}\normalsize
The operator $D'=-i\partial_{\te}\otimes\gamma+1\otimes D$ is in fact a Quillen superconnexion acting on smooth sections of the family of Hilbert spaces $H$ over $S^1$. The index theorem (\ref{toto}) computes the net number of eigenvalues of $D'$ which cross zero in a homotopy between $D'$ and $g^{-1}D'g$ \cite{CM95}. This remark really allows us to assert that (\ref{toto}) characterizes the cohomology class of the topological anomaly \cite{AG}.
\end{remark}

\vskip 1cm
%\newpage
\noindent{\bf V. Examples}\\

\noindent 1) {\it Yang-Mills anomalies:} this is a commutative example. Let $M$ be a $p$-dimensional compact Riemannian spin manifold, $p$ even. $\Ac$ is the commutative algebra $C^{\infty}(M)$ of smooth complex functions on $M$, $H$ is the Hilbert space of $L^2$-spinors on $M$, and $D$ is the usual Dirac operator. In order to construct non-trivial BRS cohomology classes, we must replace $\Ac$ by the matrix algebra $M_N(C^{\infty}(M))$ acting on $H\otimes\cc^N$, for $N$ large enough. Now $G$ is the Lie group of unitary Yang-Mills transformations $U_N(C^{\infty}(M))$.\\

Thus we are lead to compute the Chern character of the product of $(\Ac,H,D)$ by the circle. The components $\phi_n$, $n\le p+1$, of $\ch_*(-i\partial_{\te}\otimes\gamma + 1\otimes D)$ are given by (cf. \cite{CM95})
\be
\phi_n(a_0,...,a_n)= \lambda_n\int_{M\times S^1} \hat{A}(M\times S^1) \tr_N(a_0da_1...da_n)\ ,\quad a_i\in M_N(C^{\infty}(M\times S^1))\ ,
\ee
where $\hat{A}(M\times S^1)$ is the $\hat{A}$-genus of the Riemannian manifold $M\times S^1$, and $\lambda_n$ are some universal coefficients. Since $\hat{A}(M\times S^1)=\hat{A}(M)$, the local anomaly formula is a sum of terms of the form
\be
\int_{M\times S^1} \hat{A}(M) \tr_N(g^{-1}dgdg^{-1}...dg)\ ,
\ee
which is the expected result \cite{AS,Si}.\\

\noindent 2) {\it Gravitational anomalies:} this example is much more interesting, because it is highly non-commutative in nature. The algebra $\Ac$ is the crossed product $C^{\infty}(M)\cp \Gamma$, where $\Gamma$ is a pseudogroup of local diffeomorphisms of the manifold $M$. This kind of product is already used by mathematicians in the study of foliations (with $\Gamma$ discrete) \cite{C2}. The construction of $K$-cycles for the above algebra is rather involved. The difficulty is that {\it Diff$(M)$} does not preserve any Riemannian structure on $M$, so that the usual elliptic operators (Dirac, signature, etc.) cannot be defined in an invariant way. The problem is solved by working on the bundle of Riemannian metrics over $M$, on which one can construct a differential hypoelliptic operator $Q$, with invariant principal symbol \cite{CM95}. This gives rise to a spectral triple satisfying the hypothesis of the index theorem of Connes-Moscovici. The cycle is then transferred down to $M$ by Thom isomorphism.\\
Next, the Chern character of $Q$ is computed in terms of Gel'fand-Fuchs cohomology \cite{CM98}. This step requires the study of cyclic cohomology for Hopf algebras. Once transposed to our physical situation, the above construction shows that the gravitational anomaly is described by Gel'fand-Fuchs cohomology, which is more or less known to physicists. According to our presentation it can be derived from beautiful non-commutative index theorems. Details will be given elsewhere.\\

\noindent{\bf Acknowledgements:} The author wishes to thank
S. Lazzarini for numerous discussions about anomalies. He is also very
indebted to A. Connes and G. Kasparov for some remarks.

\makeatletter
\def\@biblabel#1{#1.\hfill}

\end{document}